\documentclass[twocolumn,showpacs,preprintnumbers,prl,floatfix]{revtex4-1}
\usepackage{amssymb,amsmath,amsfonts}
\usepackage{graphics,dcolumn,bm,fleqn,epic,eepic,float}
\usepackage{graphicx,epsfig} 
\usepackage{bm} 
\usepackage{subfigure}
\usepackage{color}

\begin{document}
\title{Maximal--entropy random walks in complex networks with limited information}
\author{Roberta Sinatra$^{1,2}$, Jes\'us G\'omez-Garde\~nes$^{3,4}$, 
Renaud Lambiotte$^{5}$, Vincenzo Nicosia$^{1,2}$, Vito Latora$^{1,2}$}

\affiliation{%
$^1$~Dipartimento di Fisica e Astronomia, Universit\`a di Catania and INFN, Via S. Sofia, 64, 95123 Catania, Italy}
\affiliation{%
$^2$~Laboratorio sui Sistemi Complessi, Scuola Superiore di Catania, 
Via San Nullo 5/i, 95123 Catania, Italy}
\affiliation{%
$^3$~Departamento de Fisica de la Materia Condensada, Universidad de Zaragoza, E-50009 Zaragoza, Spain}
\affiliation{$^4$~Institute for Biocomputation and Physics of Complex
Systems (BIFI), University of Zaragoza, 50009 Zaragoza, Spain}
\affiliation{%
$^5$~Institute for Mathematical Sciences, Imperial College London, 53 Prince's Gate, SW7 2PG London, UK}

%
\begin{abstract}
  {Maximization of the entropy rate is an important issue to 
  design diffusion processes aiming at a well-mixed state.} We demonstrate that 
  it is possible to construct maximal-entropy random
  walks with only \textit{local information} on the graph structure. In
  particular, we show that an almost maximal-entropy random walk is obtained
  when the step probabilities are proportional to a power of the
  degree of the target node, with an exponent $\alpha$ that depends on
  the degree-degree correlations, and is equal to 1 in uncorrelated
  graphs.
\end{abstract}
\maketitle %
{In the last decade an increasing attention has been devoted to the study of random walks on 
complex topologies \cite{bookRW,bocca}. 
Various features of random walks on networks, such as passage times \cite{rw0,rw1,fronczak} and spectral properties \cite{rw4, rw6} 
have been investigated, and random walks have also been used to detect communities \cite{rw3,rw5},
to evaluate centrality of nodes \cite{rw2,rw0,rw7} and to coarse--grain graphs \cite{rw4}. 
Another quantity recently considered is the entropy rate, a measure to characterize the mixing properties of a stochastic process \cite{cover1991}.
In particular, attention has been focused on
designing random walks with {\em maximal entropy rate} on a given
graph \cite{parry,demetrius,delvenne,burda,entropy1}, 
i.e. choosing the transition probabilities of the
random walk in such a way that the random walkers are maximally
dispersing in the graph, exploring every possible walk with equal
probability.}
{Practical examples where the maximization of entropy rate 
is important are diffusion processes which aim at well-mixing, such as spreading information about a node's state (its healthy or infected condition, its
availability or congestion, etc.) \cite{entropy1}, mixing in meta-populations models \cite{AIR}, or
global synchronization of moving agents by local entrainment \cite{diaz-guilera}.}

{In principle, the optimization of entropy rate could require 
the definition of transition probabilities relying on the 
history of the walker's positions.}
However, it has been proven that allowing a
long-term memory of the past is not needed in order to construct
maximal-entropy random walks, since it turns out that there always
exists an optimal set of transition probabilities that is Markovian
\cite{parry,demetrius,delvenne,burda}. Namely, the maximum entropy
rate is obtained with a Markov random walk in which the probability to
step from node $i$ to node $j$ is equal to $\frac{ a_{ij} u_{j} } {
  \lambda u_{i} } $, where $\lambda$ is the largest eigenvalue of the
adjacency matrix $A=\{a_{ij}\}$ of the graph, and ${\bf u}$ is the
associated eigenvector \cite{burda}.  The corresponding value of the
maximum entropy rate is equal to $\ln \lambda$.  This random walk
process has the interesting property to be biased, in the sense that a
walker follows a link $(i,j)$ with a probability proportional to the
importance of its end $j$, as measured by its eigenvector centrality
$u_j$ \cite{bonacich}. 
However, the main problem with a real
implementation is that, at each time step, the walker needs to have a
global knowledge of the network: it needs to know the adjacency matrix
of the entire graph. Such global information is very often
unavailable. A walker at a node $i$ usually has only a local
information, in the sense that it knows the first neighbors of node
$i$, and possibly some of their topological properties, such as their
degree \cite{entropy1}. 
In this
paper, we prove that almost maximal-entropy random walks can indeed be 
obtained with a limited and local knowledge of the network. We
show how to construct them by solely using the degrees of first and
second neighbors of the current node.

Let us consider a Markov random walker moving from node to node on a
connected, undirected and unweighted graph with $N$ nodes and $K$
links. At each time step, the walker moves from the current node to
one of its neighbors. If we indicate as $\pi(j|i)$ the probability of
jumping from $i$ to $j$ (with the normalization condition $\sum_j
\pi(j|i)=1 ~\forall i$), then $\pi(j|i) \neq 0$ iff $a_{ij}=1$,
i.e. if and only if $j$ belongs to the neighborhood of $i$, ${\cal
  N}_i$. We assume that the walker has the freedom to select the first
neighbors of $i$ with different probabilities, so that not all nodes
in ${\cal N}_i$ are equally selected and some of them are preferred to
the others.
The simplest case to consider is that of a {\em regular graph}, i.e. a
graph in which all nodes have the same degree. The graph can be a
random regular graph, or a regular lattice. In such a case, since all
the first neighbors of a node are equivalent, the best choice is to
set $\pi(j|i) = a_{ij} / \sum_j a_{ij}$, i.e. to select one of the
nodes in ${\cal N}_i$ at random with uniform probability. Things are
different in graphs where the nodes have different degrees, especially
in graphs with highly heterogeneous degree distributions.  As we will
show below, in these cases, maximally-random walks on a graph can be
also obtained with only a limited and local knowledge of the topology
of the graph.

The optimal random walk on a given graph can be rigorously determined
on mathematical grounds by considering the entropy rate $h$ of the
stochastic processes associated to different random walks
\cite{cover1991}.  A trajectory of $t$ steps generated by a random
walk starting at a fixed node $i$ is described by the sequence of
occupied nodes $i,i_1,i_2,\ldots,i_t$, where $i_1$,..., $i_t$ are all
indices that can take integer values between $1$ and $N$. This means
that the walker first moves from $i$ to node $i_1$, then it jumps to
node $i_2$ and so on. In practice, there is a maximum of $M(t)$
different allowed sequences of length $t$, corresponding to all
possible walks of length $t$ (and starting at node $i$) on the graph
under study.  Depending on the rules of the random walk, not all
possible sequences will appear, while some of them will occur with a
probability higher than the others. If we denote as joint probability
$p(i,i_1,i_2,\ldots,i_t)$ the probability that the sequence
$i,i_1,i_2,\ldots,i_t$ is generated by a given random walk, then the
entropy rate of the random walk, $h$, is defined as:
\begin{equation}
h = \lim_{t \rightarrow \infty} \frac{S_t}{t}\;,
\label{def_h}
\end{equation}
where $S_t$ is the Shannon entropy of the set of trajectories of
length $t$ starting at node $i$: $S_t =-\sum_{i_1,i_2,\ldots, i_t}
p(i,i_1,\ldots,i_t)\ln p(i,i_1,\ldots,i_t)$. The minimum possible
value of the entropy rate, $h_{\text{min}}=0$, is obtained when, for
large time $t$, only one trajectory dominates.  On the other hand, the
maximum possible value is obtained when, for large time $t$, all the
$M(t)$ allowed trajectories have equal probability to occur, i.e.
$p(i,i_1,\ldots,i_t) = 1/M(t)$ if $i,i_1,\ldots,i_t$ is a walk on the
graph originating in $i$, and $p(i,i_1,\ldots,i_t)=0$ otherwise. The
maximum value of the entropy is equal to: $h_{\text{max}} = \lim_{t
  \rightarrow \infty} \frac{M(t)}{t}$.  Now, in the most general case,
the probability of having a sequence of $t$ nodes originating at a
given node $i$ can be written (for any $t>1$) in terms of conditional
probabilities as: $$p(i,i_1,\ldots,i_t)=p(i_1|i)p(i_2|i,i_1)\ldots
p(i_t|i,i_1,\ldots,i_{t-1}).\;$$ Summing both ends over
$i_2,i_3,\ldots,i_t$, and using the normalization conditions
$\sum_{i_t} p(i_t|i,i_1,i_2,\ldots,i_{t-1})=1$ for $t \ge 2$, we get
an expression for the conditional probability at the first step as a
function of the $t$-times joint probabilities:
\begin{equation}
 p(i_1|i) = \sum_{i_2,i_3,\ldots,i_t} p(i,i_1,\ldots,i_t)\;.
\label{tm_vs_joint}
\end{equation}
This means that, no matter how long is the memory in the random
walker, we can always describe it as a Markov random walker, provided
that we define the transition matrix of the Markov chain $\pi(i_1|i)$
in terms of the joint probabilities $p(i,i_1,\ldots,i_t)$ as in
Eq.(\ref{tm_vs_joint}). In particular, if we want to construct a
maximal-entropy random walk, we have to set $p(i,i_1,i_2,\ldots,i_t) =
1/M(t)$ iff $i,i_1,i_2,\ldots,i_t$ is a walk on the graph, and
$p(i,i_1,i_2,\ldots,i_t)=0$ otherwise. The number of walks of length
$t$ originating in $i$ can be written in terms of the adjacency matrix
as: $M(t)=\sum_{i_1,i_2,\dots,i_t}a_{ii_1}a_{i_1i_2}\ldots
a_{i_{t-1}i_t}$.  Hence, the joint probability of a trajectory
$i,i_1,i_2,\ldots,i_t$ reads:
\begin{equation}
p(i,i_1,\ldots,i_t) = \frac{a_{ii_1}a_{i_1i_2}\ldots a_{i_{t-1}i_t}}
{\sum_{i_1,i_2,\dots,i_t}a_{ii_1}a_{i_1i_2}\ldots a_{i_{t-1}i_t}}\;,
\label{cond}
\end{equation}
and the transition matrix of the Markov random walker with the maximal 
entropy is finally given by:
\begin{equation}
 \pi(i_1|i) = \lim_{t\rightarrow \infty} \frac{ a_{ii_1} \sum_{i_2}
   a_{i_1i_2} \ldots \sum_{i_t} a_{i_{t-1}i_t} } {\sum_{i_1} a_{ii_1}
   \sum_{i_2} a_{i_1i_2} \ldots \sum_{i_t} a_{i_{t-1}i_t} }\;.
\label{tm_merw}
\end{equation}
The value of the entropy rate in Eq.~(\ref{def_h}) can then be
calculated directly from matrix $\pi$, as for any ergodic Markov
chain, from \cite{cover1991}:
\begin{equation}
   h= -\sum_{i,j}\pi(j|i) \cdot w^*(i)\ln \left[ \pi(j|i) \right]\;.
\label{eq:Entropy}
\end{equation}
where $w^*(i)$ is the $i^{th}$ component of the stationary
distribution.  From Eq.~(\ref{tm_merw}) it is clear that, in the most
general case, in order for a walker at a node $i$ to select one of its
first neighbors to step on, the walker needs to know not only which
node is in ${\cal N}_i$, but also the neighborhood of first neighbors,
the neighborhood of second neighbors, and so on. In practice, the
local choice of moving from $i$ to one particular neighbor $i_1$,
depends on the whole adjacency matrix of the graph. However, as we
demonstrate below, this global information is not necessary in most of
the cases.
{\em{Uncorrelated networks.-}} Uncorrelated graphs can be described by
the degree sequence of the nodes $\{k(1), k(2), \ldots, k(N)\}$,
corresponding to a degree distribution $P_k$, since the degree of a
node does not depend on the degree of its first neighbors. In
mathematical terms, this means that the conditional probability
$P_{k'|k}$ does not depend on $k$, and can be written in terms of the
degree distribution as: $P^{\text{unc}}_{k'|k}= k' P_{k'}/ {\langle
  k\rangle}$ where the right hand side is the probability to end up in
a node of degree $k'$ by choosing an edge at random with uniform
probability. Consequently, the average degree of the neighbors of node
$j$, $k_{nn}(j) = 1/k(j) \sum_l a_{jl} k(l)$, does not depend on the
degree of $j$, $k_{nn}(j) = k_{nn} ~\forall j$, and the last two
summations in the numerator and in the denominator of
Eq.~(\ref{tm_merw}), namely $\sum_{i_{t-2}} a_{i_{t-3}i_{t-2}}
\sum_{i_{t-1}} a_{i_{t-2}i_{t-1}} k(i_{t-1}) = \sum_{i_{t-2}}
a_{i_{t-3}i_{t-2}} k(i_{t-2}) k_{nn}(i_{t-2}) $ can be written as
$k_{nn} \sum_{i_{t-2}} a_{i_{t-3}i_{t-2}} k(i_{t-2}) $. The constant
$k_{nn}$ at the numerator and at the denominator cancels out, so that
the same argument can be repeated again and again.  Finally, the
formula factorizes into:
\begin{equation}
\pi^1(i_1|i) = \frac{ a_{ii_1} k(i_1) } { \sum_{i_1} a_{ii_1} k(i_1) }\;.
\label{tm_merw_uncorrelated}
\end{equation}
where, by the symbol $\pi^1$ we mean the first order approximation to
the transition matrix $\pi$ in Eq.~(\ref{tm_merw}).  This formula
tells us that the best diffusion process on a uncorrelated graph is a
random walk whose motion is linearly biased on node degrees.  Thus, a
walker at a given node, only needs to have information on its first
neighbors and their degree. Since the degrees of different nodes are
not correlated, local information of the degree of first neighbors
is, in this case sufficient to construct the diffusion
process with maximal entropy.  Such information is ``locally available'' to the walkers,
meaning that a walker at node $i$ has complete information on the
degree of each node in its neighborhood ${\cal N}_i$. Now, it is
intuitive that a random walk choosing a node $j$ proportionally to
$k(j)$, so that all the trajectories of length 2 starting in $i$ will
occur with the same probability, will be more random than a walker
selecting uniformly the first neighbors of $i$.

Formula (\ref{tm_merw_uncorrelated}) gives theoretical grounds to the
results of Ref.\cite{entropy1}, where random walks with power law
dependence $\pi(i_1|i) \propto k^{\alpha}(i_1)$ were explored as a
function of $\alpha$ ($\alpha >0$ indicates a bias toward high-$k$
neighbors, while $\alpha <0$ means preferring low-$k$ nodes), and it
was numerically found indication of $\alpha=1$ as the best value of
$\alpha$ if the graph is uncorrelated. Of course, if all nodes have
the same degree, as in a regular graph,
the transition matrix reduces to that of an unbiased walker:
\begin{equation}
\pi^0(i_1|i) = \frac{ a_{ii_1} } { \sum_{i_1} a_{ii_1} }\;.
\label{tm_merw_zero}
\end{equation}
This is the lowest possible approximation for $\pi$ in
Eq.~(\ref{tm_merw}): in the case of no available information, each
neighbor has the same probability to be selected. The values of $h$
obtained numerically with transition matrices $\pi^0$ and $\pi^1$ in
different models of uncorrelated networks are reported in Table
\ref{table}. In agreement with our predictions, in regular lattices
and in random regular graphs, $h(\pi^0)$ is equal to the maximal
possible entropy $h_{\text{max}}=\ln \lambda$. In Erd\H{o}s-R\'enyi
(ER) random graphs not all nodes have the same degree, so that a
random walk linearly biased on degree has an entropy $h(\pi^1)$ that
is much closer to the maximum, than $h(\pi^0)$.  This effect is even
more evident in scale-free graphs, i.e. in graphs with a very
heterogeneous degree distribution.

\begin{table}\footnotesize
\begin{tabular}{|l|c|c|c||c|}
\hline & $\frac{h(\pi^0)}{h(\pi)}$ & $\frac{h(\pi^1)}{h(\pi)}$ &
$\frac{h(\pi^2)}{h(\pi)}$ & $h_{\text{max}}=h(\pi) $ \\\hline\hline
Regular lattice                    & 1.000 & 1.000 & 1.000 & 1.79 \\\hline
Random regular graph               & 1.000 & 1.000 & 1.000 & 1.79 \\\hline
ER random graph                    & 0.954 & 0.993 & 0.998 & 1.98 \\\hline
Uncorr. scale-free $\gamma=1.5$    & 0.886 & 0.992 & 0.996 & 2.36\\\hline
BA model                           & 0.825 & 0.976 & 0.996 & 2.52\\\hline
Assort. scale-free $\gamma=1.5$    & 0.876 & 0.991 & 0.999 & 2.44 \\\hline
Disassort. scale-free $\gamma=1.5$ & 0.937 & 0.990 & 0.997 & 2.18\\\hline
\hline
Regular lattice (1\% defects)   & 0.996 & 0.997 & 0.998 & 1.38 \\\hline
Regular lattice (10\% defects)  & 0.967 & 0.978 & 0.981 & 1.34 \\\hline
Regular lattice (20\% defects)  & 0.931 & 0.955 & 0.963 & 1.29 \\\hline
\hline
Internet AS  \cite{AS}         & 0.744 & 0.900 & 0.980 & 4.10\\\hline
US Airports  \cite{AIR}        & 0.879 & 0.990 & 0.997 & 3.88\\\hline
E-Mail        \cite{Mail}      & 0.881 & 0.983 & 0.997 & 3.03\\\hline 
SCN (cond-mat)\cite{SCN}       & 0.694 & 0.867 & 0.946 & 3.17 \\\hline
SCN (astro-ph) \cite{SCN}      & 0.784 & 0.941 & 0.973 & 4.41 \\\hline
PGP         \cite{PGP2}        & 0.597 & 0.92  & 0.976 & 3.75 \\\hline
\end{tabular}
\caption{The entropies of random walks with no information,
  $h(\pi^0)$, and with local information respectively on nearest,
  $h(\pi^1)$, and next-nearest neighbors, $h(\pi^2)$, are compared to
  the maximal possible entropy $h_{\text{max}}=h(\pi) = \ln \lambda$
  on different graph models with $N=500$ and average degree $\langle k
  \rangle = 6$, on $N=40 \times 40$ regular square 
    lattices with defects (see \cite{supplementary}), and on various real networks.
    \label{table} }
\end{table}

{\em{Networks with degree-degree correlations.-}} Graphs with
degree-degree correlations are described in terms of their degree
distribution $P_k$, and of a non-trivial $P_{k'|k}$. This is because
the probability that a link from a node of degree $k$ arrives at a
node of degree $k'$ does not simply factorize in terms of the degree
distribution. In such graphs the average degree of the first neighbors
of a node $j$, $k_{nn}(j)$, does depend on $k(j)$. Therefore, in
analogy with Eq.~(\ref{tm_merw_uncorrelated}) we can define a second
order approximation of Eq.~(\ref{tm_merw}):
\begin{eqnarray}
\pi^2(i_1|i)&=&\frac{a_{ii_1}\sum_{i_2}a_{i_1 i_2}k(i_2)}{
  \sum_{i_1}a_{ii_1}\sum_{i_2}a_{i_1 i_2}k(i_2)} \nonumber
\\ 
&=&\frac{
  a_{ii_1}k(i_1)k_{nn}(i_1)}{\sum_{i_1}a_{ii_1}k(i_1)k_{nn}(i_1)}\;,
\label{tm_merw_correlated}
\end{eqnarray}
describing a Markov walker that, at each time step, selects a first
neighbor, $i_1$, of the current node, with a probability proportional
to the sum of the degrees of the first neighbors of $i_1$.  This is
equivalent to make equiprobale all the walks of length 3 originating
in $i$.  In conclusion, to construct high-entropy random walks on
correlated graphs, a walker at a given node needs to know the degree
of first and second neighbors of the current node, which is still 
local information.

In Table \ref{table} we report $h(\pi^2)$ for
various models and for real networks. In models of uncorrelated graphs
$h(\pi^2)$ is not very different from $h(\pi^1)$, while in models 
of correlated graphs, in lattices with defects 
and in most of the networks from the real world
$h(\pi^2)$ is a much better approximation of $h(\pi)$ than $h(\pi^1)$.
\begin{figure}[t!]
\epsfig{file=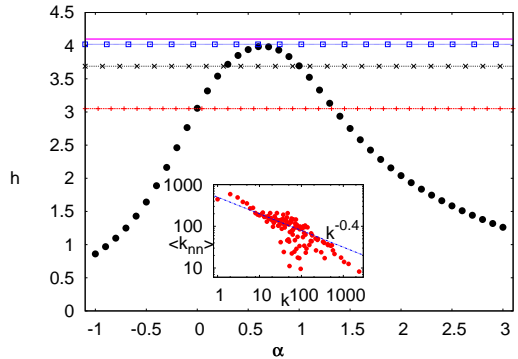,width=2.8in,angle=0,clip=1}
\caption{Entropy rate of power-law biased random walks as a function
  of the degree exponent $\alpha$ for Internet AS. Horizontal lines
  correspond to, from bottom to top, $h(\pi^0)$, $h(\pi^1)$,
  $h(\pi^2)$ and $h_{\text{max}}=h(\pi)$. (Inset) Average degree $k_{nn}$ of
  the first neighbors of nodes of degree $k$, as a function of $k$,
  with fit $k^{-0.4}$.}
\label{fig1}
\end{figure}
In most real-world networks, degree-degree correlations are such that
the average degree of the first neighbors of a node exhibits a clear
power-law dependence on degree: $k_{nn}(j) \sim [k(j)]^{-\nu}$, with
$\nu>0$ ($\nu<0$) for disassortative (assortative) networks
\cite{bocca}. For instance, as shown in the inset of
Fig.~\ref{fig1}, $\nu \simeq 0.4$ for the Internet at the autonomous
systems level \cite{AS}.  Plugging this dependence in
Eq.~(\ref{tm_merw_correlated}), we get an approximate form for the
maximal-entropy random walk in a correlated random graph in terms of
degree-biased random walks:
\begin{equation}
 \pi^2(i_1|i) \simeq    \frac{           a_{ii_1}   [k (i_1) ]^{1-\nu} } 
                    { \sum_{i_1} a_{ii_1}     [k (i_1) ]^{1-\nu} }\;.
\label{tm_merw_correlated_approx}
\end{equation}
In practice, on a correlated network, an approximation for the maximal-entropy
random walk can be obtained by considering a random walk whose motion
is biased as a power of the target node degree, with an exponent
$\alpha=1-\nu$. Hence, the optimal bias $\alpha_{opt}$ is larger
(smaller) than 1 for assortative (disassortative) networks, meaning
that we have to prefer a super-linear (sub-linear) bias on the node
degree. As an example, in Fig.~\ref{fig1} we report the entropy rate
of a biased random walk as a function of the exponent $\alpha$ on a
disassortative real-world network. We found $\alpha_{opt}=0.6$ for
Internet AS, which is perfectly in agreement with the value $\nu=0.4$
in the inset, through the relation $\alpha_{opt}=1-\nu$.  We have also
checked that this relation holds for the other real networks in Table
\ref{table}.

{\em{Networks with higher-order degree-correlations.-}} Similar
arguments can be repeated for networks with higher-order
correlations. This procedure generates a class of biased random walks
defined by the transition matrices $ \pi^0$, $ \pi^1$, $\pi^2$, etc,
incorporating more and more information about the system structure.
In Supplementary Material we studied how this sequence of transition
matrices converges to $\pi$ in different networks.  In the limit case
in which a graph has correlations at all orders, we have to rely on
the full transition matrix of Eq.~(\ref{tm_merw}), which can be also
expressed by means of the eigenvalues and eigenvectors of the
adjacency matrix of the graph.  In fact, the numerator and the
denominator of ~(\ref{tm_merw}) can be rewritten in terms of powers of
the adjacency matrix, respectively as $ a_{ii_1} \sum_{i_t}
(A^{t-1})_{i_1 i_t}= a_{ii_1} (A^{t-1}\cdot {\bf 1})_{i_1}$ and $
\sum_{i_t} (A^{t})_{i i_t}= (A^{t}\cdot {\bf 1})_{i}$, where
$(A^{t})_{i j}$ indicate the entry $i,j$ of matrix $A^t$, and ${\bf
  1}$ is a vector of ones. By making use of the power method for $t
\rightarrow \infty$, we finally get:
\begin{equation}
 \pi(i_1|i) =   \frac{  a_{ii_1}   u_{i_1} } 
                     { \lambda    u_{i}   } =  
                \frac{         a_{ii_1}   u_{i_1} } 
                     {  \sum_j a_{i j}   u_{j} }\;.
\label{tm_merw_bis}
\end{equation}
where $\lambda$ and vector ${\bf u}$ are respectively the largest eigenvalue
and its associated eigenvector of the adjacency matrix
\cite{note}.
Eq. (\ref{tm_merw_bis}) represents a Markov walk whose transition
probability is linearly biased by the components of eigenvector ${\bf u}$,
also known as the eigenvector centrality of the node \cite{bonacich},
and it is indeed the same transition matrix proposed in~\cite{burda}
as the process with the maximum possible entropy rate $h_{\text{max}}
= \ln \lambda$ \cite{parry,demetrius,delvenne,burda}.

In order to perform a maximal-entropy Markov walk on a graph, at each
time step, a walker needs a global knowledge of the whole network and
has to compute ${\bf u}$, which has $O(K)$ computational
complexity. However, global information is in practice always
unavailable in real systems. As we have shown in this paper, this
global knowledge is not necessary since in many real-world networks
long-range interactions are weak and can be neglected. It is therefore
possible to construct almost maximal-entropy random walks with only
local information on the graph structure. This can be done with $O(
\langle k \rangle)$ complexity, a dramatic improvement
which opens up to practical applications in social, biological and
technological systems.


\newpage

\section{\large{SUPPLEMENTARY MATERIAL}}

\bigskip

\normalsize
\section{Kullback--Leibler divergence}
\begin{table}[b]
  \begin{tabular}{|l||c|c|c|c|c|}
    \hline $D_{KL}(\cdot)$& $(\pi|\pi^0)$ & $(\pi|\pi^1)$ &
    $(\pi|\pi^2)$ & $(\pi|\pi^3)$ & $(\pi|\pi^4)$ \\\hline
    \hline
    Internet AS          & 0.784 & 0.163 & 0.089 & 0.032 & 0.031 \\ \hline
    US airports          & 0.928 & 0.176 & 0.072 & 0.011 & 0.001 \\ \hline
    E-mail               & 0.724 & 0.137 & 0.045 & 0.019 & 0.009 \\ \hline
    SCN (cond-mat)       & 1.796 & 0.900 & 0.737 & 0.576 & 0.471 \\ \hline
    SCN (astro)          & 2.499 & 1.167 & 0.805 & 0.570 & 0.417 \\ \hline
    PGP                  & 1.529 & 0.729 & 0.529 & 0.387 & 0.282 \\ \hline
  \end{tabular}
  \caption{Kullback--Leibler divergence between $\pi$ and successive
    approximations $\pi^k$ for different real and synthetic networks.}
  \label{table:table1}
\end{table}
In order to test the quality of the approximations of different orders
we considered the Kullback-Leibler divergence between the transition
matrix $\pi$ in Eq. (4) and the transition matrices which use only
local information.  Given two discrete distributions $P=\{p_i\}$ and
$Q=\{q_i\}$, the Kullback--Leibler divergence $D_{KL}(P|Q)$, also
known as the relative entropy of $P$ with respect to $Q$, is a
non-symmetric measure of the amount of extra information required to
represent P by using only information about Q. It is defined as the
average of the logarithmic distance between $P$ and $Q$, weighted by
the probability $P$:
 \begin{equation}
  D_{KL}(P|Q) = \sum_i p_i \log_2{\frac{p_i}{q_i}}
  \label{eq:kl}
\end{equation}
and represents the number of extra bits of information required to
reconstruct $P$ starting from $Q$~\cite{cover1991S}. In other words,
$D_{KL}$ is the expected log-likelihood to reject a false hypothesis
$Q$, per event. In this sense, $D_{KL}$ measures how much $P$ and $Q$
are different.  We have computed the Kullback--Leibler divergence
between the transition matrix $\pi$ in Eq.~(4) and the transition
matrices $\pi^0$, $\pi^1$, $\pi^2$, $\pi^3$, $\pi^4$ corresponding to
local approximations of increasing order. The expressions for $\pi^0$,
$\pi^1$, $\pi^2$ are respectively given in Eq.~(7), Eq.~(6) and
Eq.~(8), while $\pi^3$ and $\pi^4$ are defined as follows:
\begin{eqnarray}
  \pi^3(i_1|i)&=&\frac{a_{ii_1}\sum_{i_2}a_{i_1 i_2}k(i_2)k_{nn}(i_2)}{
    \sum_{i_1}a_{ii_1}\sum_{i_2}a_{i_1 i_2}k(i_2)k_{nn}(i_2)}\\
  \pi^4(i_1|i)&=&\frac{a_{ii_1}\sum_{i_2 i_3}a_{i_1 i_2}a_{i_2 i_3}k(i_3)k_{nn}(i_3)}{
    \sum_{i_1}a_{ii_1}\sum_{i_2 i_3}a_{i_1 i_2} a_{i_2 i_3}k(i_3)k_{nn}(i_3)}
\end{eqnarray}
Notice that the choice of transition matrix $\pi^k$ guarantees that
all the walks of length $k+1$ are equiprobable. Therefore, the values
of $D_{KL}(\pi|\pi^k)$ measure the inaccuracy in using the process
$\pi^k$, which makes equiprobable walks of length $k+1$, with respect
to using process $\pi$, which makes equiprobable walks of infinite
length. In Table \ref{table:table1} we report the values of
$D_{KL}(\pi|\pi^k), k=0,1,2,3,4$, obtained for the six real
networks considered in Table I of the main text.

In the first three networks in the table, $D_{KL}( \pi | \pi^2)$ is
lower than 0.1 bits. For these networks, the entropy rate $h(\pi^2)$
is about $99\%$ of the maximal entropy rate $h(\pi)$.  Conversely, for
the last three networks, the divergence $D_{KL}( \pi | \pi^2)$ is
always approximately 1 bit, and in fact the entropy rate $h(\pi^2)$ is
around $96\%$ of $h(\pi)$. As expected the divergence rapidly
decreases as we include walks of higher length.

\begin{figure}[htbp]
  \epsfig{file=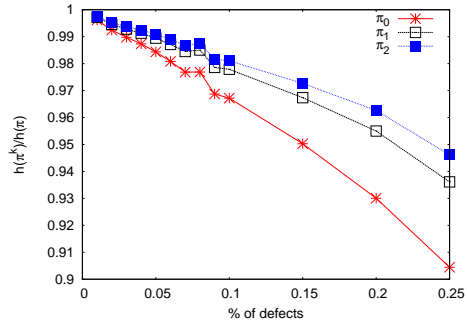,width=0.35\textwidth,angle=0,clip=1}
  \caption{\label{fig1S} Ratio of the entropy rate of approximations
    $\pi^0$ (red plus), $\pi^1$ (black squares) and $\pi^2$ (blue
    squares) over the entropy rate of a maximal-entropy random walk
    $\pi$, as a function of the percentage of links removed in
    lattices with defects, as explained in the text.}
\end{figure}

\section{Lattices with defects}
We have also studied approximations $\pi^0$, $\pi^1$ and $\pi^2$ of
the maximal-entropy random walk on lattices with some irregularities
as in Ref.~\cite{burdaS}. The networks are constructed by
removing at random a small fraction of nonadjacent links from an $L
\times L $ square lattice with periodic boundary conditions.  In this
way we obtain a damaged lattice with a weak disorder (dilution), where
most of the nodes are of degree $k=4$ and some of degree $k=3$. The
removal of links at random from a regular lattice destroys the perfect
degree regularity and introduces degree-degree correlations.  In
Fig.~\ref{fig1S} we report the ratio between the entropy rate of the
approximations $\pi^0$, $\pi^1$ and $\pi^2$, and the entropy rate of
$\pi$, as a function of the damage inflicted to the network.  Notice
that, even when $10\%$ of links have been removed (as in Ref.~\cite{burdaS}),
the approximation $\pi^2$ gives an error smaller than $2\%$. 
These numerical results confirm that local approximations provide
good estimates of the maximal entropy rate also in lattices with
defects.
  

\end{document}